\documentclass{edited_elsart}

\usepackage{natbib}

\usepackage{graphicx}
  \usepackage{lineno}

\usepackage{amsmath,amssymb}
\usepackage{macros}

\newtheorem{proposition}{Proposition}[section]

\newtheorem{definition}{Definition}[section]
\newtheorem{remark}{Remark}[section]
\newtheorem{example}{Example}[section]

\newcommand{\design}{{\mathcal D}}
\newcommand{\fraction}{{\mathcal F}}

\newenvironment{proof}{\par\noindent\emph{Proof.}}{\hfill\qed}
\begin{document}

\begin{frontmatter}



\title{Generation of Fractional Factorial Designs}

\author{Roberto Fontana\thanksref{c}}
\author{Giovanni Pistone\thanksref{c}}

\address[c]{DIMAT Politecnico di Torino, Corso Duca degli Abruzzi 24, 10129
Torino, Italy}

\begin{abstract}
The joint use of counting functions, Hilbert basis and Markov basis allows to define a procedure to generate all the fractions that satisfy a given set of constraints in terms of orthogonality. The general case of mixed level designs, without restrictions on the number of levels of each factor (like primes or power of primes) is studied. This new methodology has been experimented on some significant classes of fractional factorial designs, including mixed level orthogonal arrays. 
\end{abstract}


\end{frontmatter}

\section{Introduction}
All the fractional factorial designs that satisfy a set of conditions in terms of orthogonality between factors have been described as the zero-set of a system of polynomial equations in which the indeterminates are the complex coefficients of their counting polynomial functions (\cite{pistone|rogantin:2008-JSPI}, \cite{fontana|pistone|rogantin:2000}). A short review of this theory can be found in \cite{fontana|rogantin:sudo2008}. In Section \ref{sec:back} we report a part of it to facilitate the reader. 
In Section \ref{sec:balance} we write the problem of finding fractional factorial designs that satisfy a set of conditions as a system of linear equations in which the indeterminates are positive integers. In section \ref{sec:gof}, using 4ti2 (\cite{4ti2_1.3.1}) we find all the generators of some classes of fractional factorial designs, including mixed level orthogonal arrays and sudoku designs. Finally, in section \ref{sec:moves} we consider the moves between different fractions as integer valued functions defined over the full factorial design. We build a procedure to move between fractions that use Markov basis. 

\section{Notation and background} \label{sec:back}
\subsection{Full factorial design}

 We adopt the notation used in \cite{pistone|rogantin:2008-JSPI} and denote:
 \begin{itemize}
 \item by $\design_j$ a \emph{factor} with $n_j$ levels coded with the  $n_j$-th roots of the unity:
   \begin{equation*}\design_{j} = \{\omega_0,\ldots,\omega_{n_j-1}\}
   \qquad \omega_k=\exp\left(i\:  \frac {2\pi}{n_j} \ k\right) \ ;
   \end{equation*}
  \item by $\design$  the \emph{full factorial design} with \emph{complex coding}
 \begin{equation*}
 \design = \design_1 \times \cdots \design_j \cdots \times \design_m \ .
 \end{equation*} \
 \item by $\# \design$ the cardinality of $\design$.
\item by $L$  the \emph{full factorial design} with \emph{integer coding}
\begin{eqnarray*}
 L = \mathbb Z_{n_1} \times \cdots \times \mathbb Z_{n_j}\cdots \times  \mathbb Z_{n_m} \ ,
\end{eqnarray*}
\item
by $\alpha$ an element of $L$
$$
\alpha= (\alpha_1,\ldots,\alpha_m)  \quad \alpha_j = 0,\ldots,n_j-1, j=1,\ldots,m \ .
$$
\item by $[\alpha-\beta]$ the $m$-tuple made by the componentwise difference
 $$\left(\left[\alpha_1-\beta_1 \right]_{n_1}, \ldots, \left[\alpha_j-\beta_j \right]_{n_j}, \ldots,
\left[\alpha_m - \beta_m\right]_{n_m} \right)\ ;
$$
the computation of the $j$-th element is in the ring $\mathbb Z_{n_j}$.
 \item by $X_j$  the $j$-th component function, which maps a  point to its $i$-th component:
\begin{equation*}
X_j : \quad \design \ni (\zeta_1,\ldots,\zeta_m)\ \longmapsto \ \zeta_j \in \design_j \ ;
\end{equation*}
the function $X_j$ is called \emph{simple term} or, by abuse of terminology, \emph{factor}.
 \item by $X^\alpha$  the \emph{interaction term} $X_1^{\alpha_1} \cdots X_m^{\alpha_m}$, i.e. the function
\begin{equation*}
X^\alpha : \quad \design \ni (\zeta_1,\ldots,\zeta_m)\ \mapsto \ \zeta_1^{\alpha_1}\cdots \zeta_m^{\alpha_m} \ ;
\end{equation*}
 \end{itemize}

We notice that $L$ is both the full factorial design with integer coding and \emph{the exponent set of all the simple
factors and interaction terms} and  $\alpha$ is both a treatment combination in the integer coding and a multi-exponent
of an interaction term.

The full factorial design in complex coding is identified as the zero-set in $\mathbb C^m$ of the system of polynomial
equations
\begin{equation}\label{eq:design}
X_j^{n_j}-1=0 \quad , \qquad  j=1,\ldots,m \ .
\end{equation}

\begin{definition}
\begin{enumerate}
\item A \emph{response} $f$ on the design $\design$ is a $\mathbb C$-valued polynomial function defined on $\design$.
\item
The \emph{mean value} on $\design$ of a response $f$, denoted by $E_{\design}(f) $, is:
\begin{equation*}
E_{\design}(f)  = \frac 1 {\#\design} \sum_{\zeta \in \design} f(\zeta) \ .
\end{equation*}
\item
A response $f$ is \emph{centered} on $\design$  if $E_\design(f) = 0$. Two responses $f$ and $g$ are \emph{orthogonal
on $\design$} if $E_\design(f \ \overline g) = 0$, where $\overline g$ is the complex conjugate of $g$.
\end{enumerate}
\end{definition}
It should be noticed that the set of all the responses is a complex Hilbert space with the Hermitian product:
 $$ f\cdot g = E_\design(f \ \overline g) \ . $$
Moreover
\begin{enumerate}
 \item $X^\alpha\overline{X^\beta}=X^{[\alpha-\beta]}$;
 \item $E_{\design}(X^0)=1$, and $E_{\design}(X^\alpha)=0$ for $\alpha
 \neq 0$.
\end{enumerate}

The set of functions $\left\{X^\alpha \ , \ \alpha \in L \right \}$ is an orthonormal basis of the \emph{complex
responses} on  design $\design$. In fact $\#L=\#\design$ and, from  properties (i) and (ii) above, it follows that:
\begin{equation*}E_{\design}(X^\alpha\overline{X^\beta}) = E_{\design}(X^{[\alpha-\beta]}) =
\begin{cases}
1 & \text{if }  \alpha=\beta \\
0 & \text{if }  \alpha \neq \beta
\end{cases}\end{equation*}

In particular,  each response $f$ can  be represented as a unique $\mathbb C$-linear combination of constant, simple
and interaction terms. This representation is obtained by repeated applications of the re-writing rules derived from
Equations (\ref{eq:design}).  Such a polynomial is called the {\em normal form} of $f$ on $\design$. In this paper we intend that all the computation are made using the normal form.

\begin{example} \label{ex:2to3}

Consider the $2^3$ full factorial design. All the monomial responses  on $\design$ are $$ 1, \ X_1, \ X_2 , \ X_3, \
X_1 X_2 , \ X_1 X_3 , \ X_2 X_3 , \ X_1 X_2 X_3 $$ or, equivalently,
$$X^{(0,0,0)},  X^{(1,0,0)}, X^{(0,1,0)}, X^{(0,0,1)}, X^{(1,1,0)}, X^{(1,0,1)}, X^{(0,1,1)}, X^{(1,1,1)}
$$
and $L$ is
$$ L=\{(0,0,0),(1,0,0),(0,1,0),(0,0,1),(1,1,0),(1,0,1),(0,1,1),(1,1,1)\} \ .
$$
\end{example}

\subsection{Fractions of a full factorial design}
A fraction $\fraction$ is a multiset $(\fraction_*,f_*)$ whose underlying set of elements $\fraction_*$ is contained in $\design$ and $f_*$ is the multiplicity function $f_*: \fraction_* \rightarrow \mathbb N$ that for each element in $\fraction_*$ gives the number of times it belongs to the multiset $\fraction$. 

All fractions can be obtained by
adding polynomial equations, called \emph{generating equations}  to the design equations \ref{eq:design}, in order to
restrict the number of solutions.

\begin{definition}

If $f$ is a response on $\design$ then its \emph{mean value on  $\fraction$}, denoted by $E_{\fraction}(f)$, is
$$E_{\fraction}(f)= \frac 1 {\# \fraction} \sum_{\zeta\in\fraction} f(\zeta)$$ where $\# \fraction$ is the total number
of treatment  combinations of the fraction.

A response $f$ is \emph{centered} if $E_\fraction(f) = 0$. Two responses $f$ and $g$ are \emph{orthogonal on
$\fraction$} if $E_\fraction(f \ \overline g) = 0$.
\end{definition}

With the complex coding the vector orthogonality of two interaction terms $X^\alpha$ and $X^\beta$ as defined before
(with respect to a given Hermitian product) corresponds to the combinatorial orthogonality (all the level combinations
appear equally often in $X^\alpha X^\beta$).

We consider the general case in which fractions can contain points that are replicated.

\begin{definition} \label{de:indicator}

The \emph{counting function} $R$ of a fraction $\fraction$ is a response defined on $\design$ so that for each $\zeta \in \design$, $R(\zeta)$ equals the number of appearances of $\zeta$ in the fraction. A $0-1$ valued counting function is called \emph{indicator function} of a single replicate fraction $\fraction$.
We denote by $c_\alpha$ the coefficients of the representation of $R$  on $\design$ using  the monomial basis
$\{X^\alpha, \ \alpha \in L\}$:
\begin{equation*}
R(\zeta) = \sum_{\alpha \in L} c_\alpha X^\alpha(\zeta) \quad \zeta\in\design \quad  c_\alpha \in \mathbb C \ .
\end{equation*}
\end{definition}

As the counting function is real valued, we have $\overline{c_\alpha} = c_{[-\alpha]}$. We will write $c_0$ in place of $c_{0,\dots,0}$.

\begin{remark}
The counting function $R$ coincides with multiplicity function $f_*$.
\end{remark}

\begin{proposition} \label{pr:bc-alpha}

Let $\fraction$ be a fraction of a full factorial design $\design$ and $R = \sum_{\alpha \in L} c_\alpha X^\alpha$ be
its counting function.
\begin{enumerate}
 \item \label{it:balpha}
The coefficients $c_\alpha$ are:
 $$c_\alpha= \frac 1 {\#\design} \sum_{\zeta \in \fraction} \overline{X^\alpha(\zeta)} \ ;$$
in particular, $c_0$ is the ratio between the number of points of the fraction and that of the design.
 \item \label{it:system}
In a fraction without replications, the coefficients $c_\alpha$ are related according to: $$ c_\alpha = \sum_{\beta \in L} c_\beta \ c_{[\alpha - \beta]} \
.$$
 \item The term $X^\alpha$ is centered on $\fraction$, i.e. $\mathbb E_{\fraction}(X^\alpha)$, if, and only if,
 $$c_\alpha=c_{[-\alpha]}=0 \ . $$
 \item The terms $X^\alpha$ and $X^\beta$
are orthogonal on $\fraction$, i.e. $\mathbb E_{\fraction}(X^\alpha \ \overline{X^\beta})=0$, if, and only if,
$$c_{[\alpha-\beta]}=0 \ .$$
\end{enumerate}
\end{proposition}

\begin{example}

We consider the fraction $\fraction = \{(-1,-1,1), (-1,1,-1)\}$ of the $2^3$ full factorial design of Example
\ref{ex:2to3}. All the monomial responses on $\fraction$ and their values on the points are
\begin{equation*}
\begin{array}{c|r|r|r|r|r|r|r|r}
 \zeta &1 & X_1 & X_2 & X_3 & X_1 X_2 & X_1 X_3 & X_2 X_3& X_1 X_2 X_3 \\ \hline
(-1,-1,1) & 1 &  -1 &  -1 &   1 &       1 &       -1 &      -1 &       1  \\
 (-1,1,-1) & 1 &  -1 &   1 &  -1 &      -1 &       1 &      -1 &       1
\end{array}
\end{equation*}
Using Item \ref{it:balpha} of Proposition \ref{pr:bc-alpha}, it is easy to compute the coefficients $c_\alpha$: \
 $ c_{(0,1,0)}= c_{(0,0,1)}= c_{(1,1,0)}=c_{(1,0,1)}=0$; \
 $ c_{(0,0,0)}=c_{(1,1,1)}= \frac 2 4$ and $c_{(1,0,0)}= c_{(0,1,1)}= - \frac 2 4 $.
Hence, the indicator function is $$ F= \frac 1 2 \left( 1 - X_1 -X_2 X_3+X_1X_2 X_3\right) \ .$$ From the null
coefficients we see that $X_1$ and $X_3$ are centered and that $X_1$ is orthogonal to both $X_2$ and $X_3$. \hfill
$\square$
\end{example}

\subsection{Projectivity and orthogonal arrays}

\begin{definition}

A fraction $\fraction$ {\em factorially projects} onto the $I$-factors, $I\subset \{1,\ldots,m\}$, if the projection is
a multiple full factorial design, i.e. a full factorial design where each point appears equally often. A fraction $\fraction$ is a {\em mixed orthogonal
array} of strength $t$ if it factorially projects onto any $I$-factors with $\#I=t$.
\end{definition}
Strength $t$ means that, for any choice of $t$ columns of the matrix design, all possible combinations of symbols
appear equally often.

\begin{proposition} [Projectivity] \label{pr:projectivity}
\begin{enumerate}
\item \label{it:fact} A fraction \emph{factorially projects onto the $I$-factors} if, and only if,
 all the coefficients of the counting function involving only the $I$-factors are 0.
 \item If there exists
a subset $J$ of $\{1,\ldots,m\}$ such that the $J$-factors appear in all the non null elements of the counting
function,
 the fraction \emph{factorially projects onto the $I$-factors}, with $I=J^c$.
 \item A fraction is an \emph{orthogonal array of strength $t$}  if, and
 only if, all the coefficients of the counting function up to the order $t$ are zero:
 $$ c_{\alpha}=0 \quad \textrm{for all} \ \alpha  \textrm{ of order up to }t, \ \alpha \ne (0,0, \ldots ,0)\ .
$$
\end{enumerate}
\end{proposition}

\begin{example} [Orthogonal array]

The fraction of a $2^5$ full factorial design {\footnotesize
\begin{eqnarray*} \fraction_O&=& \{(-1, -1, -1, -1, -1, 1), ( -1, -1, -1, 1, 1, 1), ( -1, -1, 1, -1, -1, -1),
\\ & & ( -1, -1, 1, 1, 1, -1),( -1, 1, -1, -1, -1, -1), ( -1, 1, -1, 1, 1, -1), ( -1, 1, 1, -1, 1, 1),
\\ & & ( -1, 1, 1, 1, -1,1 ), ( 1, -1, -1, -1, 1, 1), ( 1, -1, -1, 1, -1, 1), ( 1, -1, 1, -1, 1, -1),
\\ & & ( 1, -1, 1, 1, -1, -1),( 1, 1, -1, -1, 1, -1), ( 1, 1, -1, 1, -1, -1), ( 1, 1, 1, -1, -1, 1),
\\ & &( 1, 1, 1, 1, 1, 1)\}
\end{eqnarray*}}
 is an orthogonal array of strength 2; in fact, its indicator function
 \begin{eqnarray*}
  F &=&\frac 1  4+ \frac 1  4 X_2X_3X_6- \frac 1 8 X_1X_4X_5+ \frac 1 8 X_1X_4X_5X_6+ \frac 1 8 X_1X_3X_4X_5\\
 & &
 + \frac 1 8 X_1X_2X_4X_5+ \frac 1 8 X_1X_3X_4X_5X_6+ \frac 1 8 X_1X_2X_4X_5X_6
 \\
 & &+ \frac 1 8 X_1X_2X_3X_4X_5-\frac 1 8 X_1X_2X_3X_4X_5X_6
  \end{eqnarray*}
 contains only terms of order greater than 2, together with the constant term. \hfill $\square$
\end{example}

\section{Counting functions and strata} \label{sec:balance}
%In this paper we will work with $m$ factors, each one with the same number of level $s$, with $s$ prime. The full design $\design$ is $(\{\omega_0,\ldots,\omega_{s-1}\})^m$, while $L$ is $\mathbb Z_s^m$ .

From Proposition \ref{pr:bc-alpha} and Proposition \ref{pr:projectivity} we have that the problem of finding fractional factorial designs that satisfy a set of conditions in terms of orthogonality between factors can be written as a polynomial system in which the indeterminates are the complex coefficients $c_\alpha$ of the counting polynomial fraction.

\begin{example}
Let's consider 3 factors, each one with two levels. The indicator functions $F=\sum_\alpha {c_\alpha X^{\alpha}}$ such that the terms $X_1, X_2, X_3$ are centered on $\fraction$ and the terms $X_i, X_j$ $i,j=1,2,3, i\neq j$ are orthogonal on $\fraction$, where $\fraction = \{\zeta \in \design: F(\zeta)=1 \}$, are those for which the following conditions on the coefficients of $F$ holds
\begin{equation*}
\begin{cases}
c_0=c_0^2 + c_{123}^2 \\
c_{123}=2 c_0 c_{123} 
\end{cases}\end{equation*}

Apart from the trivial $F=0$, i.e. $\fraction=\emptyset$ and $F=1$, i.e. $\fraction=\design$ we find 
$F=\frac{1}{2} ( 1+X_1 X_2 X_3)$ and $F=\frac{1}{2} ( 1-X_1 X_2 X_3)$
\end{example}

Let's now introduce a different way to describe the full factorial design $\design$ and all its subsets. Let's  consider the indicator functions $1_{\zeta}$ of all the single points of $\design$
\begin{equation*}
1_{\zeta} : \quad \design \ni (\zeta_1,\ldots,\zeta_m)\ \mapsto \begin{cases}
1 \quad \zeta=(\zeta_1,\ldots,\zeta_m) \\
0 \quad \zeta \neq (\zeta_1,\ldots,\zeta_m)
\end{cases}
\end{equation*}
It follows that the counting function $R$ of a fraction $\fraction$ can be written as
\[
\sum_{\zeta \in \design} y_{\zeta} 1_{\zeta}
\]  
with $y_{\zeta} \equiv R(\zeta) \in \{0,1,\ldots,n,\ldots\}$. The particular case in which $R$ is an indicator function corresponds to 
$y_{\zeta} \in \{0,1\}$.

The coefficients $y_{\zeta}$ are related to the coefficients $c_\alpha$ as in the following Proposition \ref{pr:beqy}

\begin{proposition} \label{pr:beqy}
Let $\fraction$ be a fraction of $\design$. Its counting fraction $R$ can be expressed both as $R=\sum_\alpha c_\alpha X^\alpha$ and $R=\sum_{\zeta \in \design} y_{\zeta} 1_{\zeta}$. The relation between the coefficients $c_\alpha$ and $y_{\zeta}$ is
\[
c_\alpha=\frac{1}{\#\design}\sum_{\zeta \in \design} y_{\zeta} \overline{X^\alpha(\zeta)} 
\]
\end{proposition}
\begin{proof}
From Proposition \ref{pr:bc-alpha} we have
\begin{eqnarray*}
c_\alpha&=&\frac{1}{\#\design}\sum_{\zeta \in \fraction} \overline{X^\alpha(\zeta)} = \\
%&=& \frac{1}{\#\design}\sum_{t \in \design} R(t) \overline{X^\alpha(t)} = \\
%&=& \frac{1}{\#\design}\sum_{t \in \design} \left( \sum_{\zeta \in \design} y_{\zeta} 1_{\zeta}(t)\right) \overline{X^\alpha(t)} = \\
%&=& \frac{1}{\#\design}\sum_{\zeta \in \design} y_{\zeta} \left(\sum_{t \in \design} 1_{\zeta}(t) \overline{X^\alpha(t)}\right) = \\
&=& \frac{1}{\#\design}\sum_{\zeta \in \design} y_{\zeta}  \overline{X^\alpha(\zeta)}
\end{eqnarray*}
\end{proof}

\subsection{Strata}
As described in Section \ref{sec:back}, we consider $m$ factors, $\design_1, \ldots, \design_m$ where
$\design_j \equiv \Omega_{n_j}=\{\omega_0, \ldots, \omega_{n_j-1}\}$, for $j=1,\ldots,m$. From \cite{pistone|rogantin:2008-JSPI}, we recall two basic properties which hold true for the full design $\design$

\begin{proposition} \label{pr:P-1}
Let $X_j$ the simple term with level set $\Omega_{n_j} = \{\omega_0, \ldots, \omega_{n_j-1}\}$. Let's consider the term $X_j^r$ and let's define 
\[
s_j=
\begin{cases} 
1 & \; r=0 \\
n_j/gcd(r,n_j) &  \; r>0 
\end{cases}
\]
Over $\design$, the term $X_j^r$ takes all the values of $\Omega_{s_j}$ equally often.  
\end{proposition} 

\begin{proposition} \label{pr:P-2}
Let $X^\alpha=X^{\alpha_1}_{1}\cdots X^{\alpha_m}_{m}$ an interaction. $X^{\alpha_i}_i$ takes values in $\Omega_{s_i}$ where $s_i$ is determined according to the previous Proposition \ref{pr:P-1}. Let's define $s=lcm(s_1,\ldots,s_m)$. Over $\design$, the term $X^\alpha$ takes all the values of $\Omega_s$ equally often.  
\end{proposition} 

Let's now define the strata that are associated to simple and interaction terms.

\begin{definition}
Given a term $X^\alpha, \alpha \in L=\mathbb Z_{n_1} \times \ldots \times \Z_{n_m}$ the full design $\design$ is partitioned into the the following strata
\[
D_h^\alpha = \left \{ \zeta \in \design: \overline{X^\alpha(\zeta)} = \omega_h \right \}
\]
where $\omega_h \in \Omega_s$ and $s$ is determined according to the previous Propositions \ref{pr:P-1} and \ref{pr:P-2}. 
\end{definition}

\begin{remark}
We define strata using the conjugate $\overline{X^\alpha}$ of the term in place of the term ${X^\alpha}$ itself because it will simplify the notations.  
\end{remark}

\begin{remark}
Each stratum is a regular fraction whose defining equation is $X^\alpha(\zeta)=\omega_{-h}$, \cite{pistone|rogantin:2008-JSPI}.
\end{remark}

We use $n_{\alpha,h}$ to denote the number of points of the fraction $\fraction$ that are in the stratum $D_h^\alpha$, with $h=0,\dots,s-1$,
\[
n_{\alpha,h}=\sum_{\zeta \in D_h^\alpha} y_\zeta
\]

The following Proposition \ref{pr:ceqn} links the coefficients $c_\alpha$ with $n_{\alpha,h}$.
\begin{proposition} \label{pr:ceqn}
Let $\fraction$ be a fraction of $\design$ with counting fraction $R=\sum_{\alpha \in L} c_\alpha X^\alpha$. Each $c_\alpha, \alpha \in L$, depends on $n_{\alpha,h}, h=0,\ldots,s-1$,  as
\[
c_\alpha=\frac{1}{\#\design} \sum_{h=0}^{s-1} n_{\alpha,h} \omega_h
\]
where $s$ is determined by $X^\alpha$ (see Proposition \ref{pr:P-2}).
Viceversa, each $n_{\alpha,h}, h=0,\ldots,s-1$, depends on $c_{[-k\alpha]}, k=0,\ldots,s-1$ as  
\[
n_{\alpha,h}=\frac{\#\design}{s} \sum_{k=0}^{s-1} c_{[-k \alpha]} \omega_{[hk]}
\]
\end{proposition}
\begin{proof}
Using Proposition \ref{pr:beqy}, it follows that we can write the coefficients $c_\alpha$ in the following way
\[
c_\alpha =  \frac{1}{\#\design}\sum_{\zeta \in \design} y_{\zeta} \overline{X^\alpha(\zeta)} =\frac{1}{\#\design} \sum_{h=0}^{s-1} \omega_h \sum_{\zeta \in D_h^\alpha} y_{\zeta} = \frac{1}{\#\design} \sum_{h=0}^{s-1} n_{\alpha,h} \omega_h
\]
For the viceversa, we observe the indicator function of strata can be obtained as follows.
We define 
\[
\tilde{F}_{0}^s(\zeta)=\sum_{k=0}^{s-1} \zeta^k = 
\begin{cases}
\frac{1-\zeta^s}{1-\zeta} & \text{if } \zeta \neq 1 \\
s & \text{if } \zeta=1
\end{cases}
\] 
We have $\tilde{F}_{0}^s(\omega_k)=0$ for all $\omega_k \in \Omega_s, k \neq 0$. It follows that
\[
F_{\alpha,0}(\zeta) = \frac{1}{s} \tilde{F}_{0}^s(\zeta^\alpha)= \frac{1}{s} \left( 1+\zeta^\alpha+\ldots+\zeta^{(s-1)\alpha} \right)
\] 
is the indicator function associated to $D_0^\alpha$.

The indicator of $D_h^\alpha=\left \{ \zeta \in \design: \overline{X^\alpha(\zeta)} = \omega_h \right \}=
\left \{ \zeta \in \design: X^\alpha(\zeta) = \omega_{[-h]} \right \}$ will be 
\[
F_{\alpha,h}(\zeta)=F_{0}^s(\omega_{h} \zeta^\alpha)= \frac{1}{s} \left( 1+\omega_{h}\zeta^\alpha+\ldots+\omega_{[(s-1)h]}\zeta^{(s-1)\alpha}\right )
\]
We get
\begin{eqnarray*}
n_{\alpha,h} &=& \sum_{\zeta \in D_h^\alpha} R(\zeta) = \sum_{\zeta \in \design} F_{\alpha,h}(\zeta)R(\zeta)=\\
&=& \sum{\zeta \in \design} \left( \frac{1}{s} \sum_{k=0}^{s-1} \omega_{[kh]} X^{k\alpha}(\zeta) \right) \left( \sum_{\beta} c_\beta X^{\beta}(\zeta)\right)= \\
&=& \frac{\#\design}{s} \sum_{k,\beta: [k\alpha+\beta]=0} \omega_{[kh]} c_\beta = \frac{\#\design}{s} \sum_{k=0}^{s-1} \omega_{[kh]} c_{[-k\alpha]} \\
\end{eqnarray*}

\end{proof}

\begin{remark}
From Proposition \ref{pr:ceqn} we get
\begin{eqnarray*}
n_{0,h}&=&0, \; h=1,\ldots,s-1 \\
n_{\alpha,0}&=&\frac{\#\design}{s}\sum_{k=0}^{s-1} c_{[-k\alpha]}
\end{eqnarray*}
and in particular $n_{0,0}=\#\fraction$.
\end{remark}
We now use a part of Proposition 3 of \cite{pistone|rogantin:2008-JSPI} to get conditions on $n_{\alpha,h}$ that makes $X^\alpha$ centered on the fraction $\fraction$.

\begin{proposition} \label{pr:pr3jspi}
Let $X^\alpha$ be a term with level set $\Omega_s$ on full design $\design$. Let $P(\zeta)$ the complex polynomial associated to the sequence $(n_{\alpha,h})_{h=0,\ldots,s-1}$ so that
\[
P(\zeta)= \sum_{h=0}^{s-1} n_{\alpha,h} \zeta^{h}
\]
and let's denote by $\Phi_s$ the cyclotomic polynomial of the $s$-roots of the unity.
\begin{enumerate}
\item Let $s$ be prime. The term $X^\alpha$ is centered on the fraction $\fraction$ if, and only if, its $s$ levels appear equally often:
\[
n_{\alpha,0}=n_{\alpha,1}=\ldots=n_{\alpha,s-1}=\lambda_\alpha
\] 
\item Let $s=p_1^{h_1} \dots p_d^{h_d}$ with $p_i$ prime, for $i=1,\ldots,d$. The term $X^\alpha$ is centered on the fraction $\fraction$ if, and only if, the remainder
\[
H(\zeta)=P(\zeta) \text{ mod } \Phi_s(\zeta)
\]
whose coefficients are integer linear combinations of $n_{\alpha,h}, h=0,\ldots,s-1$, is identically zero. 
\end{enumerate}
\end{proposition}

\begin{proof}
See Proposition 3 of \cite{pistone|rogantin:2008-JSPI}. 
\end{proof}

\begin{remark}
Being $D_h^\alpha$ a partition of $\design$, if $s$ is prime we get $\lambda_\alpha=\frac{\#\fraction}{s}$.
\end{remark}
%The following Lemma \ref{le:roots} provides a useful result for the linear combinations of the $s$-th roots of the unity.

%\begin{lemma} \label{le:roots}
%Given $r_0,\ldots,r_{s-1} \in \mathbb R$, the relation
%\[
%r_0\omega_0+r_1\omega_1+\ldots+r_{s-1}\omega_{s-1}=0 
%\]
%is equivalent to
%\[
%r_0=r_1=\ldots=r_{s-1} 
%\]
%\end{lemma}
%\begin{proof}
%If $r_0=r_1=\ldots=r_{s-1}$ the proof follows from the well-known relationship between the $s$-th roots of the %unity  
%\[
%\omega_0+\omega_1+\ldots+\omega_{s-1}=0
%\]
%For the viceversa we refer to \cite{pistone|rogantin:2008-JSPI} and to \cite{lang:65}.
%Without loss of generality, we suppose $r_{s-1} \neq 0$ and we define the following polynomial
%\[
%P(X)=\sum_{i=0}^{s-1} {r_i X^i}
%\]
%Let's take one of the $s$-th roots of the unity, $\omega_h$, and let's divide $P(X)$ by $X-\omega_h$.
%The remainder is
%\[
%r_0+r_{s-1}\omega_{s-1}  + \ldots +r_1 \omega_1 
%\]
%By hypothesis, $r_0\omega_0+r_1\omega_1+\ldots+r_{s-1}\omega_{s-1}=0$ and so $P(\omega_h)=0$ for each $h \in \{0,\ldots,s-1\}$.
%
%Being $s$ a prime number, the cyclotomic polynomial is
%\[
%\Phi_s(X)=\sum_{k=0}^{s-1} {X^k}
%\]
%The polynomial $P$ is divided by the cyclotomic polynomial and both $P$ and $\Phi_s$ have the same degree. It should be  $P(X)=r_{s-1} \Phi_s(X)$ and therefore 
%\[
%r_0=r_1=\ldots=r_{s-1} 
%\]
%\end{proof}

If we remind that $n_{\alpha,h}$ are related to the values of the counting function $R$ of a fraction $\fraction$ by the following relation
\[
n_{\alpha,h}=\sum_{\zeta \in D_h^\alpha} y_\zeta,
\]
this Proposition \ref{pr:pr3jspi} allows to express the condition \emph{$X^\alpha$ is centered on $\fraction$} as integer linear combinations of the values $R(\zeta)$ of the counting function over the full design $\design$. In the Section \ref{sec:gof}, we will show the use of this property to generate fractional factorial designs.
  
%We can express constraints on the coefficients $c_\alpha$ of a counting function $R=\sum_{\alpha} c_{\alpha} X^{\alpha}$ in terms of the values $y_{\zeta}$ that the counting function $R$ takes on the points $\zeta \in \design$.
 
%\begin{proposition} \label{pr:c-strata}
%Given a counting function $R=\sum_{\alpha} c_{\alpha} X^{\alpha} = \sum_{\zeta \in \design} y_{\zeta} 1_{\zeta}$, %the following conditions are equivalent:
%\begin{enumerate}
%\item[(i)]$c_\alpha=0$ 
%\item [(ii)] $n_{\alpha,0} = n_{\alpha,1} = \ldots = n_{\alpha,s-1}$
%\item [(iii)] $n_{\alpha,h} = \frac{\#\fraction}{s}, \; h=0,\ldots, s-1$
%\end{enumerate} 
%\end{proposition} 
%\begin{proof}

%(ii) $\Leftrightarrow$ (i)

%Using Proposition \ref{pr:ceqn}, it follows that we can write the coefficients $c_\alpha$ as
%\[
%c_\alpha =   \frac{1}{\#\design} \sum_{h=0}^{s-1} n_{\alpha,h} \omega_h
%\]
%For $\alpha=0$ we get $c_0 =  \frac{1}{\#\design}  \sum_{p \in \design} y_p$

%Now, by Lemma \ref{le:roots}, we complete this part of the proof.

%(ii) $\Rightarrow$ (iii)

%$\{ D_h^\alpha, h=0,\ldots,s-1 \}$ is a partition of $\design$ so $\sum_{h=0}^{s-1} n_{\alpha,h} = \#\fraction$. If $n_{\alpha,h}, \; h=0,\ldots,s-1$ are all equal it follows $n_{\alpha,h} = \frac{\#\fraction}{s}, \; h=0,\ldots, s-1$. 

%(iii) $\Rightarrow$ (ii) 

%It is straightforward.
%\end{proof}

We conclude this section limiting to the particular case where all factors have the same number of levels $s$ and $s$ is prime. We provide some results concerning the coefficients of counting functions, regular fractions, wordlength patterns and margins. 

\subsection{Coefficients of the polynomial counting function}
From Proposition \ref{pr:pr3jspi} we get  the following result on the coefficients of a counting function
\begin{proposition} \label{pr:samestrata}
Given a counting function $R=\sum_{\alpha} c_{\alpha} X^{\alpha}$, if $c_\alpha=0$ then $c_{[k\cdot \alpha]}=0$ for all $k=1,\ldots,s-1$, where $[k\cdot \alpha]$ is $\underbrace{\alpha+\ldots+\alpha}_{k \; times}$ in the ring $\mathbb Z_{s}^m$.
\end{proposition}
\begin{proof}
Let's consider $c_{k\cdot \alpha}$. From Proposition \ref{pr:pr3jspi}, $c_{k\cdot \alpha}$ is equal to zero if, and only if, 
\[
\sum_{\zeta \in D_0^{k\cdot \alpha}} y_{\zeta} = \sum_{\zeta \in D_1^{k\cdot \alpha}} y_{\zeta} = \ldots = \sum_{\zeta \in D_{s-1}^{k\cdot \alpha}} y_{\zeta}
\]
We observe that 
\begin{eqnarray*}
D_h^{k\cdot \alpha} &=& \left \{ \zeta \in \design: \overline{X^{k\cdot \alpha}(\zeta)} = \omega_h \right \} = \\
&=& \left \{ \zeta \in \design: \overline{X^{\alpha}(\zeta)}^k = \omega_h \right \}=
\left \{ \zeta \in \design: \overline{X^{\alpha}(\zeta)}= \omega_{[kh]} \right \} =
D_{[kh]}^{\alpha}
\end{eqnarray*}
where $[kh]$ is $\underbrace{h+\ldots+h}_{k \; times}$ in the ring $\mathbb Z_{s}$.

It follows that $X^{\alpha}$ and $X^{k \cdot \alpha}$ partition $\design$ in the same strata and therefore we get the proof.
\end{proof}

\subsection{Regular designs}
Let's consider a fraction $\fraction$ without replicates and with indicator function $F=\sum_{\alpha}c_\alpha X^\alpha$. Proposition 5 in (\cite{pistone|rogantin:2008-JSPI}) states  that a fraction $\fraction$ is regular if, and only if, its indicator function $F$ has the form
\[
F=\frac{1}{l}\sum_{\alpha \in \mathcal L} \overline{e(\alpha)}X^\alpha
\]
where $\mathcal L \subseteq L$, $\mathcal L$ is a subgroup of $L$ and $e:\mathcal L \rightarrow \{\omega_0,\ldots,\omega_{s-1}\}$ is a given mapping.

If we use Proposition \ref{pr:pr3jspi} we immediately get a characterisation of regular fractions based on the frequencies $n_{\alpha,h}$.

\begin{proposition}
Given a single replicate fraction $\fraction$ with indicator function $F=\sum_{\alpha} c_{\alpha}X^\alpha$ the following statements are equivalent:
\begin{enumerate}
\item[(i)] $\fraction$ is regular
\item [(ii)] for $n_{\alpha,h}$ there are only two possibilities
\begin{enumerate}
 \item if $c_\alpha=0$ then $n_{\alpha,h} = \frac{\#\fraction}{s}, \; h=0,\ldots,s-1$,
 \item if $c_\alpha \neq 0$ then $\exists h_{*} \in \{0,\ldots,s-1\}$ such that 
 \begin{equation*}
 n_{\alpha,h}= 
 \begin{cases}
\frac{\#\design}{l} & \text{if } h=h_{*}\\
0 & \text{otherwise}
 \end{cases}
  \end{equation*}
\end{enumerate}
\end{enumerate}
\end{proposition} 
\begin{proof}
Using Proposition \ref{pr:ceqn} we get
\[
c_\alpha =  \frac{1}{\#\design} \sum_{h=0}^{s-1} n_{\alpha,h} \omega_h
\]
Proposition 5 in \cite{{pistone|rogantin:2008-JSPI}} gives the following conditions on the  coefficients of the indicator function $F$ of a regular fraction $\fraction$:
\[
c_\alpha=\begin{cases}
\frac{\overline{e(\alpha)}}{l}, & \alpha \in \mathcal L \subseteq L \\
0 & \text{otherwise}
\end{cases}
\]
where $e:\mathcal L \rightarrow \{\omega_0,\ldots,\omega_{s-1}\}$, $l=\#mathcal L$ and $\mathcal L$ is a subgroup of $L$.

Let's consider $\alpha \in \mathcal L$. We get
\[
\frac{1}{\#\design} \sum_{h=0}^{s-1} n_{\alpha,h} \omega_h = \frac{\overline{e(\alpha)}}{l}
\]
Let's suppose $e(\alpha)=\omega_{h_*}$. We obtain
\begin{equation} \label{eq}
\frac{1}{\#\design}  \sum_{h=0, h\neq h_*}^{s-1} n_{\alpha,h} \omega_h + (\frac{1}{\#\design} n_{\alpha,h_*} - \frac{1}{l}) \omega_{h_*}   = 0
\end{equation}
To simplify the notation we let $a_h=\frac{1}{\#\design}n_{\alpha,h}, h=0,\ldots,s-1, h \neq h_*$ and $a_{h_*}=\frac{1}{\#\design} n_{\alpha,h_*} - \frac{1}{l}$. Therefore, from the proof of item (1) of Proposition \ref{pr:pr3jspi}, for the relation \ref{eq} to be valid, it should be
\[
a_0 = a_1 = \ldots = a_{s-1}
\]
Being $\sum_{h=0}^{s-1} n_{\alpha,h} = \#\fraction$ it follows 
\[
\sum_{h=0}^{s-1} n_{\alpha,h}=\sum_{h=0, h\neq h_*}^{s-1} (\#\design) a_h +   (\#\design)( a_{h_*} + \frac{1}{l}) = (\#\design) \sum_{h=0}^{s-1} a_{h} + \frac{(\#\design)}{l} = \#\fraction
\]
and so
\[
a_h = \frac{1}{s(\#\design)} (\#\fraction - \frac{(\#\design)}{l} )
\]
We finally get
\[
n_{\alpha,h} =
\begin{cases}
\frac{1}{s} (\#\fraction - \frac{(\#\design)}{l} ) + \frac{(\#\design)}{l} & \text{if } h=h_*\\
\frac{1}{s} (\#\fraction - \frac{(\#\design)}{l} ) & \text{ otherwise} 
\end{cases}
\]
Being $\mathcal L$ a subgroup of $L$ it follows that $0 \in \mathcal L$ and so $c_0=1/l$. We also know that $c_0=\frac{\#\fraction}{\#\design}$ and therefore
\[ 
\#\fraction=\frac{\#\design}{l} 
\]
For the null coefficients of $F$, $\{c_\alpha: \alpha \in L - \mathcal L \}$, it is enough to use Proposition \ref{pr:c-strata} to conclude the proof.
\end{proof}
%\begin{remark}
%From this Proposition, being $n_{0,0}=\#\fraction$ and $n_{0,h}=0, h=1,\ldots,s-1$ we easily derive a condition for the number of non null coefficient of the indicator function $F$ of a regular fraction:
%\[
%l=\frac{\#\design}{\#\fraction}=\frac{1}{c_0}
%]
%\end{remark}

\subsection{Wordlength Pattern} \label{sec:wlp}
Aberration is often used as a criterion to compare fractional factorial designs. The generalized minimum aberration, proposed by \cite{xu|wu:2001}, is based on the generalised wordlength pattern, see also \cite{beder|willenbring:2009}. 
It can be shown that the generalized wordlengths can be written in terms of the squares of the modules of the coefficients $c_\alpha$, obtaining
\[
A_j = \left( \frac{\#\design}{\#\fraction} \right)^2 \sum_{wt(\alpha)=j} {\left| c_\alpha \right|^2} = \frac{1}{c_0^2} \sum_{wt(\alpha)=j} {\left| c_\alpha \right|^2} \; \text{ for } j=1,\ldots,m
\]
where $wt(\alpha)$ is the Hamming weight of $\alpha$, i.e. the number of nonzero components of $\alpha$.
We now express the square of the module of the coefficient $c_\alpha$ in terms of $n_{\alpha,h}$.
\begin{proposition} \label{pr:modulo}
\begin{equation*}
\left|  c_\alpha \right|^2  = \frac{1}{(\#\design)^2} \sum_{h=0}^{s-1} ( n_{\alpha,h}^2 - n_{\alpha,h} n_{[\alpha,h-\gamma]}) \; \text{ for } \gamma \in \{1,\ldots,s-1\} 
\end{equation*}
\end{proposition}
\begin{proof}
From Proposition \ref{pr:ceqn} we get
\[
c_\alpha =  \frac{1}{\#\design} \sum_{h=0}^{s-1} n_{\alpha,h} \omega_h
\]
It follows 
\begin{eqnarray*} \label{eq1}
\left| c_\alpha \right|^2 &=& c_\alpha \overline{c_\alpha} = \\
                 &=&\frac{1}{(\#\design)^2}( \sum_{h=0}^{s-1} n_{\alpha,h} \omega_h)( \sum_{k=0}^{s-1} n_{\alpha,k} \overline{\omega_k}) = \\
                 &=& \frac{1}{(\#\design)^2} ( \sum_{h=0}^{s-1} n_{\alpha,h} \omega_h) ( \sum_{k=0}^{s-1} n_{\alpha,k} \omega_{[s-k]}) = \\
                 &=& \frac{1}{(\#\design)^2} \sum_{\gamma=0}^{s-1} \sum_{p=0}^{s-1} n_{\alpha,p} n_{[\alpha,p-\gamma]} \omega_\gamma 
\end{eqnarray*}

$\left| c_\alpha \right|^2$ must be a real number. Being $\omega_0=1$ it follows
\begin{eqnarray} \label{eq1}
(\frac{1}{(\#\design)^2} \sum_{p=0}^{s-1} n_{\alpha,p}^2 - \left| c_\alpha \right|^2)\omega_0 +  \frac{1}{(\#\design)^2} \sum_{\gamma=1}^{s-1} \sum_{p=0}^{s-1} n_{\alpha,p} n_{[\alpha,p-\gamma]} \omega_\gamma = 0
\end{eqnarray}
To simplify the notation we let $a_0=(\frac{1}{(\#\design)^2} \sum_{p=0}^{s-1} n_{\alpha,p}^2 - \left| c_\alpha \right|^2)$ and $a_\gamma=\frac{1}{(\#\design)^2} \sum_{p=0}^{s-1} n_{\alpha,p} n_{[\alpha,p-\gamma]}, \gamma=1,\ldots,s-1$. Therefore, by Lemma \ref{le:roots}, for the relation \ref{eq1} to be valid, it should be
\[
a_0 = a_1 = \ldots = a_{s-1}
\]
Using one of the equalities, $a_0=a_h$ $h=1,\ldots,s-1$, it follows
\[
\left| c_\alpha \right|^2 = \frac{1}{(\#\design)^2} \sum_{p=0}^{s-1} ( n_{\alpha,p}^2 - n_{\alpha,p} n_{[\alpha,p-h]})
\]
\end{proof}

\begin{remark}
Proposition \ref{pr:modulo} provides a useful tool to compute the modules of the coefficients $c_\alpha$. Indeed it is enough to choose $\gamma=1$ and compute $\left| c_\alpha \right|^2$ as $\frac{1}{(\#\design)^2} \sum_{h=0}^{s-1} ( n_{\alpha,h}^2 - n_{\alpha,h} n_{[\alpha,h-1]})$;
\end{remark}

\begin{remark}
We make explicit these relations for $2$ and $3$ level fraction.

If $s=2$ then 
\[
\left| c_\alpha \right|^2  = \frac{1}{(\#\design)^2} ( n_{\alpha,0} -  n_{\alpha,1})^2
\]
If $s=3$ then, choosing $\gamma=1$, 
\[
\left| c_\alpha \right|^2  = \frac{1}{(\#\design)^2} ( n_{\alpha,0}^2+ n_{\alpha,1}^2 + n_{\alpha,2}^2 - 
n_{\alpha,0} n_{\alpha,2} - n_{\alpha,1} n_{\alpha,0} - n_{\alpha,2} n_{\alpha,1})
\]
\end{remark}

\begin{remark}
We observe that, denoting by $\overline{n}_\alpha$ the mean of the values of $n_{\alpha,h}$, $\overline{n}_\alpha=\frac{1}{s} \sum_{h=0}^{s-1} n_{\alpha,h}$, we get
\[
\sum_{h=0}^{s-1} {(n_{\alpha,h} - \overline{n}_\alpha)^2} = \sum_{h=0}^{s-1} {n_{\alpha,h}^2} - s \overline{n}_\alpha^2
\]
We have
\begin{eqnarray*}
\overline{n}_\alpha^2 &=& \frac{1}{s^2} \sum_{h,k=0}^{s-1} n_{\alpha,h} n_{\alpha,k} = \\
&=& \frac{1}{s^2} \left( \sum_{h=0}^{s-1} {n_{\alpha,h}^2} + 2 \sum_{h=0}^{s-1} n_{\alpha,h} n_{\alpha,[h-1]} + \ldots 2 \sum_{h=0}^{s-1} n_{\alpha,h} n_{\alpha,[h-s_*]} \right)
\end{eqnarray*}
where $s_*=\frac{s-1}{2}$.
Proposition \ref{pr:modulo} states that all the quantities $\sum_{h=0}^{s-1} n_{\alpha,h} n_{\alpha,[h-\gamma]}$ are equal and so, choosing, without loss of generality, $\gamma=1$, we get
\[
\overline{n}_\alpha^2 = \frac{1}{s^2}  \left( \sum_{h=0}^{s-1} {n_{\alpha,h}^2} + 2 s_* \sum_{h=0}^{s-1} n_{\alpha,h} n_{\alpha,[h-1]} \right) = \frac{1}{s^2}  \left( \sum_{h=0}^{s-1} {n_{\alpha,h}^2} + (s-1) \sum_{h=0}^{s-1} n_{\alpha,h} n_{\alpha,[h-1]} \right)
\]
and therefore 
\begin{eqnarray*}
\sum_{h=0}^{s-1} {(n_{\alpha,h} - \overline{n}_\alpha)^2} &=& \sum_{h=0}^{s-1} {n_{\alpha,h}^2} - s \overline{n}_\alpha^2 = \\
&=& \frac{s-1}{s} \left( \sum_{h=0}^{s-1} {n_{\alpha,h}^2} - \sum_{h=0}^{s-1} n_{\alpha,h} n_{\alpha,[h-1]} \right) = \\
&=& \frac{s-1}{s} (\# \design)^2 \left| c_\alpha \right|^2
\end{eqnarray*}
It follows that, if we denote by $\sigma_\alpha^2$ the variance of $n_{\alpha,h}$, $\sigma_\alpha^2=\frac{1}{s}  \sum_{h=0}^{s-1} {(n_{\alpha,h} - \overline{n}_\alpha)^2}$ we get
\[
\left|  c_\alpha \right|^2 = \left( \frac{s^2}{(s-1) (\# \design)^2} \right) \sigma_\alpha^2  
\]
and so the square of the module of $c_\alpha$ represents, apart from a multiplicative constant, the variance of the frequencies $n_{\alpha,h}$.  

\end{remark}
\subsection{Margins} \label{sec:margin}

We now examine the relationship between the margins and the coefficients of the counting functions. We refer to (\cite{pistone|rogantin:2008-JSPI}) and we report here a part of it. 

For each point $\zeta \in \design$ we consider the decomposition $\zeta=(\zeta_I,\zeta_J)$ where $I \subseteq \{1,\dots,m\}$ and $J = \{1,\dots,m\} - I \equiv I^c$ is its complement. We denote by $R_I(\zeta_I)$  the number of points in $\fraction$ whose projection on the $I$ factors is $\zeta_I$.

In particular if $I=\{1,\ldots,m\}$ we have $R_I=R$ and if $I = \emptyset$ we have $R_I=\# \fraction$. 

We denote by $L_I$ the subset of the exponents restricted to the $I$ factors and by $\alpha_I$ an
element of $L_I$:
\[
L_I=\{a_I =(\alpha_1,\dots,\alpha_m), \alpha_j=0 \text{ if } j\in J\}
\]
Then for each $\alpha \in L$  and $\zeta \in \design$ we have $\alpha=\alpha_I+\alpha_J$ and $X^\alpha(\zeta)=X^\alpha_I(\zeta_I) X^\alpha_j(\zeta_J)$. Finally we denote by $\design_I$ and $\design_J$ the full factorial over the $I$ factors and $J$ factors, respectively ($\design=\design_I \times \design_J$).

We have the following proposition (see item 1 and 2 of Proposition 4 of \cite{pistone|rogantin:2008-JSPI})

\begin{proposition} \label{pr:pr4_gpmpr}
Given a fraction $\fraction$ of $\design$
\begin{enumerate}
\item the number of replicates of the points of $\fraction$ projected on the $I$ factors is:
\[
R_I(\zeta_I)=\#\design_J\sum_{\alpha_I}c_{\alpha_I} X^{\alpha_I}(\zeta_I)
\]
\item $\fraction$ fully projects on the $I$ factors if, and only if,
\[
R_I(\zeta_I)=\#\design_J \cdot c_0 = \#\design_J \frac{\#\fraction}{\#\design} = \frac{\#\fraction}{\#\design_I}
\]
\end{enumerate}
\end{proposition}

We will refer to $R_I$ as $k$-margin, where $k=\#I$. The number of $k$-margins is $\binom{m}{k}$ and each $k$-margin can be computed over $s^{k}$ points $\zeta_I \in \design_I$. It follows that there are $(1+s)^m$ marginal values in total. 

Using item 1 of Proposition \ref{pr:pr4_gpmpr} and reminding that we work with a prime number of level s we have
\[
R_I(\zeta_I)=s^{m-k} \sum_{\alpha_I} c_{\alpha_I} \zeta_I^{\alpha_I}
\]
or, by the definition of $R_I$ as the restriction of $R$ over the $I$ factors,
\[
\sum_{\zeta_J \in \design_J} R(\zeta_I,\zeta_J) \equiv \sum_{\zeta_J \in \design_J} y_{\zeta_I,\zeta_J} = s^{m-k} \sum_{\alpha_I} c_{\alpha_I} \zeta_I^{\alpha_I}
\]

We point out the following relationship between margins.
\begin{proposition} \label{pr:hierarchy}
If $A \subseteq B \subseteq \{1,\dots,m\}$ and $R_B(\zeta_B)=s^{m-k_B} c_0$ then $R_A(\zeta_A)=s^{m-k_A} c_{0}$ where $\#B=k_B$ and $\#A=k_A$
\end{proposition}
\begin{proof}
Let's put $A_1=B - A$. We have
\[
R_A(\zeta_A)=\sum_{\zeta_{A_1}\in A_1} R_{A \cup A_1}(\zeta_A,\zeta_{A_1})= \sum_{\zeta_{A_1}\in A_1} R_B(\zeta_A,\zeta_{A_1})= s^{k_B-k_A} s^{m-k_B} c_{0}= s^{m-k_A} c_{0}
\]
\end{proof}

We finally observe that, as we already pointed out, given $\mathcal C \subseteq L$ a set of conditions $c_\alpha=0, \alpha \in \mathcal C$ translates in a set of conditions $\sum_{\zeta \in D_h^\alpha} y_\zeta = \lambda, h=0,\dots,s-1, \alpha \in \mathcal C$ where $\lambda$ does not depend by $\alpha$ (and by $h$).
In general, with respect to margins, the situation is different. For example let's suppose to have a $\fraction$ that fully projects over the $I_1$ and the $I_2$ factors, with $I_1 \cap I_2= \emptyset$ and $\#I_1 \neq \#I_2$.
From Proposition \ref{pr:pr4_gpmpr} we obtain 
\[
R_{I_1}(\zeta_{I_1})=\frac{\#\design}{s^{\#I_1}} \text{ and }
R_{I_2}(\zeta_{I_2})=\frac{\#\design}{s^{\#I_2}} 
\]

\section{Generation of fractions} \label{sec:gof}
Let use strata to generate fractions that satisfy a given set of constrains on the coefficients of their counting functions. Formally we give the following definition
\begin{definition}
A counting function $R=\sum_{\alpha} {c_\alpha X^{\alpha}}$ associated to $\fraction$ is a $\mathcal{C}$-compatible counting function if its coefficients satisfy to 
\[
c_{\alpha}=0, \; \alpha \in \mathcal{C}, \; \mathcal{C} \subseteq \mathbb Z_{n_1} \times \ldots \mathbb Z_{n_m}
\] 
\end{definition}
We will denote by $OF(n_1 \dots n_m,\mathcal C)$ the set of all the fractions whose counting functions are $\mathcal C$-compatible.

In the next sections, we will show our methodology on Orthogonal Arrays and Sudoku designs.

\subsection{$OA(n,s^m,t)$}
Let's consider $OA(n,s^m,t)$, i.e. orthogonal arrays with $n$ rows and $m$ columns where each columns has $s$ symbols, $s$ prime and with strength $t$. 

Using Proposition \ref{pr:projectivity} we have that the coefficients of the corresponding counting functions must satisfy the conditions $c_\alpha=0$ for all $\alpha \in \mathcal{C}$ where $\mathcal{C} \subseteq L = \{ \alpha : 0 < \|\alpha \| \leq t \}$ where $\|\alpha \|$ is the number of non null elements of $\alpha$. We have $N_1=\sum_{k=1}^t \binom{m}{k} (s-1)^k$ coefficients that must be null. 

It follows that $OF(s^m,\mathcal C) =\bigcup_{n} OA(n,s^m,t)$.
 
Now using Proposition \ref{pr:pr3jspi}, we can express these conditions using strata. If we consider $\alpha \in \mathcal C$ we write the condition $c_\alpha=0$ as    
\begin{equation*} 
\begin{cases}
\sum_{\zeta \in D_0^\alpha} y_\zeta = \lambda\\
\sum_{\zeta \in D_1^\alpha} y_\zeta = \lambda\\
\dots \\
\sum_{\zeta \in D_{s-1}^\alpha} y_\zeta = \lambda
\end{cases}
\end{equation*}

To obtain all the conditions it is enough to vary $\alpha \in \mathcal{C}$. We use Proposition \ref{pr:samestrata} to limit to the $\alpha$ that give different strata. It is easy to show that we obtain $N_2 = \frac{N_1}{s-1}$ different $\alpha$, each of them generate $s$ linear equations, for a total of \[
N=s N_2 = s \sum_{k=1}^t \binom{m}{k} (s-1)^{k-1}
\] 
constraints on the values of the counting function over $\design$.

We therefore get the following system of linear equations 
\[
A Y = \lambda \underline{1}
\]
where $A$ is the $(N \times s^m)$ matrix whose rows contains the values, over $\design$, of the indicator function of the strata, $1_{D_h^\alpha}$, $Y$ is the $s^m$ column vector whose entries are the values of the counting function over $\design$, $\lambda$ will be equal to  $\frac{\#\fraction}{s}$ and $\underline{1}$ is the $s^m$ column vector whose entries are all equal to $1$. 
We can write an equivalent  homogeneous system if we consider $\lambda$ as a new variable. We obtain
\[
\tilde{A} \tilde{Y} = 0
\]
where 
\[
\tilde{A} = \left[ 
\begin{array}{c|c}
	A & \begin{array}{r}
			-1 \\
			-1 \\
			\dots \\
			-1
\end{array}
\end{array}
\right] = \left[ A ,-\underline{1} \right]
\]
and
\[
\tilde{Y} = \left[ 
\begin{array}{c}
Y \\
\hline
\lambda
\end{array}
\right] = \left( Y , \lambda \right)
\] 

In an equivalent way, we can also express the conditions $c_\alpha=0$ for all $\alpha \in \mathcal{C}$ in terms of margins. We obtain
\[
R_I(\zeta_I) = s^{m-(\#I)} c_{0}
\]
where $I \subseteq \{1,\dots,m \}$ and $ 1 \leq \#I \leq t$. If we recall Proposition \ref{pr:hierarchy}, we can limit to the margins $R_I$ where $\#I=t$. We have $s^t \binom{m}{t}$ values of such $t$ margin
\[
\sum_{\zeta_J \in \design_J} y_{\zeta_I,\zeta_J} = s^{m-t} c_{0}
\]
In this case, with the same approach that we adopted for strata, we obtain a system of linear equations
\[
B Y = \rho \underline{1}
\]
where $\rho=s^{m-t} c_{0}$ and its equivalent homogeneous system
\[
\tilde{B} \tilde{Y} = 0
\]
Now we can find all the generators of $OF(s^m,\mathcal C$, that means of Orthogonal Arrays $OA(n,s^m,t)$, by computing the Hilbert Basis corresponding to $\tilde{A}$ (or, equivalently, to $\tilde{B})$. This approach is the same of \cite{carlini|pistone:2007} but, in that work, the following conditions were used
\[
c_\alpha= \frac 1 {\#\design} \sum_{\zeta \in \fraction} \overline{X^\alpha(\zeta)} =\frac 1 {\#\design} \sum_{\zeta \in \design}  \overline{X^\alpha(\zeta)} y_\zeta =0
\]
The advantage of using strata (or margins) is that we avoid computations with complex numbers ($\overline{X^\alpha(\zeta)}$). We explain this point in a couple of examples. For the computation we use 4ti2 (\cite{4ti2_1.3.1}). 

We use both $\tilde{A}$ (strata) and $\tilde{B}$ (margins) because, even if they are fully equivalent from the point of view of the solutions that they generate, they perform differently from the point of view of the computational speed.

\subsubsection {$OA(n,2^5,2)$} \label{subsec:oa2_5}
$OA(n,2^5,2)$ were investigated in \cite{carlini|pistone:2007}. We build both the matrix $\tilde{A}$ and $\tilde{B}$. They have $30$ rows and $40$ rows, respectively and $33$ columns. 
%The matrices are available in the Annex. 
We find the same $26,142$ solutions as in the cited paper. 

\subsubsection {$OA(n,3^3,2)$} \label{subsec:oa3_3}
We build both the matrix $\tilde{A}$ and $\tilde{B}$. They have $54$ rows and $27$ rows, respectively and $28$ columns. 
%The matrices are available in the Annex. 
We find $66$ solutions, $12$ have $9$ points, all different and $54$ have $18$ points, $17$ different. 

Finally we point out that 4ti2 allows to specify upper bounds for variables. For example, if we use $\tilde{B}$ and we are interested in single replicate orthogonal arrays, we can set $1$ as the upper bound for $y_{\zeta}, \zeta \in \design$. The upper bound for the variable $\rho$ can be set to $s^{m-t} \equiv 3^{3-2}$ that corresponds to $c_0=1$, i.e. to the full design $\design$.

\subsection{$OA(n,n_1 \dots n_m,t)$}
Let's now consider the general case in which we do not put restrictions on the number of levels.

\subsubsection {$OA(n,4^2,1)$} \label{subsec:oa4_2}
In this case the number of levels is a power of a prime, $2^2$. Using Proposition \ref{pr:projectivity} we have that the coefficients of the corresponding counting functions must satisfy the conditions $c_\alpha=0$ for all $\alpha \in \mathcal{C}$ where $\mathcal{C} \subseteq L = \{ \alpha : \|\alpha \| =1 \}$.

Let's consider $c_{1,0}$. From Proposition \ref{pr:P-1} we have that $X_1$ takes the values in $\Omega_s$ where $s=4$. From Proposition \ref{pr:pr3jspi}, $X_1$ will be centered on $\fraction$ if, and only if, the remainder
\[
H(\zeta)= P(\zeta) \text{ mod } \Phi_4(\zeta)
\]  
is identically zero.
We have $\Phi_4(\zeta)=1+\zeta^2$ (see \cite{lang:65}) and so we can compute the remainder
\[
H(\zeta)=n_{(1,0),0} - n_{(1,0),2} +  (n_{(1,0),1} - n_{(1,0),3}) \zeta
\]
The condition $H(\zeta)$ identically zero translates into
\[
\begin{cases}
n_{(1,0),0} - n_{(1,0),2}=0 \\
n_{(1,0),1} - n_{(1,0),3}=0
\end{cases}
\]
Let's now consider $c_{2,0}$. From Proposition \ref{pr:P-1} we have that $X_1^2$ takes the values in $\Omega_s$ where $s=2$. From Proposition \ref{pr:pr3jspi}, $X_1^2$ will be centered on $\fraction$ if, and only if, the remainder
\[
H(\zeta)= P(\zeta) \text{ mod } \Phi_2(\zeta)
\]  
is identically zero.
We have $\Phi_2(\zeta)=1+\zeta$ (see \cite{lang:65}) and so we can compute the remainder
\[
H(\zeta)=n_{(2,0),0} - n_{(2,0),1}
\]
If we repeat the same procedure for all the $\alpha$ such that $\|\alpha \| =1$ and we recall that 
\[
n_{\alpha,h}=\sum_{\zeta \in D_h^\alpha} y_\zeta
\]
orthogonal arrays $OA(n,4^2,1)$ become the integer solutions of the following integer linear homogeneous system
%\begin{table}
%\centering
%\begin{tabular}{r r r r r r r r r r r r r r r r}
\[
\left[
\begin{array}{r r r r r r r r r r r r r r r r}
1 &  0 &  -1 &  0 &  1 &  0 &  -1 &  0 &  1 &  0 &  -1 &  0 &  1 &  0 &  -1 &  0 \\
0 &  1 &  0 &  -1 &  0 &  1 &  0 &  -1 &  0 &  1 &  0 &  -1 &  0 &  1 &  0 &  -1 \\
1 &  -1 &  1 &  -1 &  1 &  -1 &  1 &  -1 &  1 &  -1 &  1 &  -1 &  1 &  -1 &  1 &  -1 \\
1 &  0 &  -1 &  0 &  1 &  0 &  -1 &  0 &  1 &  0 &  -1 &  0 &  1 &  0 &  -1 &  0 \\
0 &  -1 &  0 &  1 &  0 &  -1 &  0 &  1 &  0 &  -1 &  0 &  1 &  0 &  -1 &  0 &  1 \\
1 &  1 &  1 &  1 &  0 &  0 &  0 &  0 &  -1 &  -1 &  -1 &  -1 &  0 &  0 &  0 &  0 \\
0 &  0 &  0 &  0 &  1 &  1 &  1 &  1 &  0 &  0 &  0 &  0 &  -1 &  -1 &  -1 &  -1 \\
1 &  1 &  1 &  1 &  -1 &  -1 &  -1 &  -1 &  1 &  1 &  1 &  1 &  -1 &  -1 &  -1 &  -1 \\
1 &  1 &  1 &  1 &  0 &  0 &  0 &  0 &  -1 &  -1 &  -1 &  -1 &  0 &  0 &  0 &  0 \\
0 &  0 &  0 &  0 &  -1 &  -1 &  -1 &  -1 &  0 &  0 &  0 &  0 &  1 &  1 &  1 &  1
\end{array}
\right]
\left[
\begin{array}{c}
y_{00} \\
y_{10} \\
y_{20} \\
y_{30} \\
y_{01} \\
y_{11} \\
y_{21} \\
y_{31} \\
y_{02} \\
y_{12} \\
y_{22} \\
y_{32} \\
y_{03} \\
y_{13} \\
y_{23} \\
y_{33} 
\end{array}
\right]
\]
%\end{tabular}
%\end{table}
Using 4ti2 we find $24$ solutions that correspond to all the Latin Hypercupe Designs (LHD).

\subsubsection {$OA(n,6^2,1)$} \label{subsec:oa6_2}
As in the previous examples,  using Proposition \ref{pr:projectivity} we have that the coefficients of the corresponding counting functions must satisfy the conditions $c_\alpha=0$ for all $\alpha \in \mathcal{C}$ where $\mathcal{C} \subseteq L = \{ \alpha : \|\alpha \| =1 \}$.

Let's consider $c_{1,0}$. From Proposition \ref{pr:P-1} we have that $X_1$ takes the values in $\Omega_s$ where $s=6$. From Proposition \ref{pr:pr3jspi}, $X_1$ will be centered on $\fraction$ if, and only if, the remainder
\[
H(\zeta)= P(\zeta) \text{ mod } \Phi_6(\zeta)
\]  
is identically zero.
We have $\Phi_6(\zeta)=1-\zeta+\zeta^2$ (see \cite{lang:65}) and so we can compute the remainder
\[
H(\zeta)=n_{(1,0),0} - n_{(1,0),2} - n_{(1,0),3} + n_{(1,0),6}+  (n_{(1,0),1} + n_{(1,0),2} - n_{(1,0),5} - n_{(1,0),6}) \zeta
\]
If we repeat the same procedure for all the $\alpha$ such that $\|\alpha \| =1$ and we recall that 
\[
n_{\alpha,h}=\sum_{\zeta \in D_h^\alpha} y_\zeta
\]
orthogonal arrays $OA(n,6^2,1)$ become the integer solutions of an integer linear homogeneous system $A R=0$ where the matrix $A$ is built as in the previous case of $OA(n,4^2,1)$.
Using 4ti2 we find $620$ solutions that correspond to all the Latin Hypercupe Designs (LHD).

% INSERIRE SUDO 
 \subsection{Sudoku designs}
As shown in \cite{fontana|rogantin:sudo2008}, a sudoku can be described using its indicator function. Here we report a very short synthesis of Section 1.3 of that work. 

A $p^2 \times p^2$ with $p$ prime sudoku design can be seen as a fraction $\fraction$ of the full factorial design $\design$:
$$
\design = R_1 \times R_2 \times C_1 \times C_2 \times S_1 \times S_2
$$
where each factor is coded with the $p$-th roots of the unity. $R_1$ and $R_2$, $C_1$ and $C_2$, $S_1$ and $S_2$, represent the rows, the columns and the symbols of the sudoku grid, respectively.

The following proposition (Proposition 5 of \cite{fontana|rogantin:sudo2008}) holds.
\begin{proposition} \label{pr:pr5}
Let $F$ be the indicator function of a fraction $\fraction$ of a design $design$, $F=\sum_{\alpha \in L} b_\alpha
X^\alpha$. The fraction $\fraction$ corresponds to a sudoku grid if and only if the coefficients $b_\alpha$ satisfy the
following conditions:
\begin{enumerate}
\item \label{it:b0}
$b_{000000} =  1/{p^2}$, i.e. the ratio between the number of points of the fraction and the number of  points of the
full factorial design is $ 1/{p^2}$;
\item \label{it:b}
for all $i_j \in \left\{0,1,\dots,p-1\right\}$:
\begin{enumerate}
\item $b_{i_1 i_2 i_3 i_4 0 0}=0$ for $(i_1, i_2, i_3, i_4) \neq (0,0,0,0)$,
\item $b_{i_1 i_2 0 0 i_5 i_6}=0$ for $(i_1, i_2, i_5, i_6) \neq (0,0,0,0)$,
\item $b_{0 0 i_3 i_4 i_5 i_6}=0$ for $(i_3, i_4, i_5, i_6) \neq (0,0,0,0)$,
\item $b_{i_1 0 i_3 0 i_5 i_6}=0$ for $(i_1, i_3, i_5, i_6) \neq (0,0,0,0)$
\end{enumerate}
i.e. the fraction factorially projects onto the first four factors and onto both symbol factors and row/column/box
factors, respectively.
\end{enumerate}
\end{proposition}
 
From this Proposition, we define $\mathcal C$ as the union of $\mathcal C_1$, $\mathcal C_2$, $\mathcal C_3$ and $\mathcal C_4$, where
\begin{eqnarray*}
\mathcal C_1 &=&\{ (i_1 i_2 i_3 i_4 0 0): (i_1, i_2, i_3, i_4) \neq (0,0,0,0)\} \\
\mathcal C_2 &=&\{ (i_1 i_2 0 0 i_5 i_6): (i_1, i_2, i_5, i_6) \neq (0,0,0,0)\} \\
\mathcal C_3 &=&\{ (0 0 i_3 i_4 i_5 i_6): (i_3, i_4, i_5, i_6) \neq (0,0,0,0)\} \\
\mathcal C_4 &=&\{ (i_1 0 i_3 0 i_5 i_6): (i_1, i_3, i_5, i_6) \neq (0,0,0,0)\} \\
\end{eqnarray*}
The problem of finding Sudoku becomes equivalent to find $\mathcal C$-compatible counting functions, that are (i) indicator functions and (ii) that satisfy the additional requirement $b_0=1/{p^2}$.  

\subsubsection {$4 \times 4$ Sudoku} \label{subsec:sudo4_4}
We use the conditions $\mathcal C$ to build both the matrices $\tilde{A}$ and $\tilde{B}$. $\tilde{A}$ has $78$ rows. With respect to $\tilde{B}$, that corresponds to the margins that must be constant, if we recall Proposition \ref{pr:hierarchy} we obtain $64$ constraints, all corresponding to $4$-margins. 
%Both the matrices are given in the Annex. 

To find all sudoku we use 4ti2, specifying the upper bounds for all the $65$ variables. The upper bounds for $y_\zeta, \zeta \in \design$ must be equal to $1$. If we use $\tilde{A}$,  the upper bound for $\lambda$ must be set equal to $\frac{\#\fraction}{s}\equiv \frac{16}{2}=8$, while if we use $\tilde{b}$ the upper bound for $\rho$ must be set equal to $s^{m-k}b_0 \equiv 2^2 \frac{1}{4}=1$.

We find all the $288$ different $4 \times 4$ sudoku as in \cite{fontana|rogantin:sudo2008}. 
We point out that to solve the problem using $\tilde{A}$ the total time was $31.59$ minutes, while using $\tilde{B}$ the total time was only $58.04$ seconds on the same computer.

If we admit counting functions with values in $\{0,1,2\}$ and $\#\fraction \leq 32$ we find $55,992$ solutions.
 
\section{Moves} \label{sec:moves}
Sometimes, given a set of conditions $\mathcal C$ we are interested in picking up a solution more than in finding all the generators. The basic idea is to generate somehow a starting solution and then to randomly walk in the set of all the solutions for a certain number of steps, taking the arrival point as a new but still $\mathcal C$-compatible counting function.

Let's use the previous results on strata to get a suitable set of \emph{moves}. We will show this procedure in the case in which all the factors have the same number of levels $s$, $S$ prime, but it can also be applied to the general case.  In Section \ref{sec:gof} we have shown that counting functions must satisfy the following set of linear equations
\[
A Y = \lambda \underline{1}
\]
where $A$ corresponds to the set of conditions $\mathcal C$ written in terms of strata.

It follows that if, given a $\mathcal C$-compatible solution $Y$, such that $A Y= \lambda \underline{1}$, we search for an additive move $X$ such that $A(Y+X)$ is still equal to $\lambda \underline{1}$, we have to solve the following linear homogenous system
\[
A X = 0
\]   
with $X=(x_\zeta), \zeta \in \design$, $x_\zeta \in \mathbb Z$ and $y_\zeta+x_\zeta \geq 0$ for all $\zeta \in \design$. We observe that this set of conditions allows to determine new $\mathcal C$-compatible solutions \emph{that give the same $\lambda$}. We know that $\lambda=\frac{\#\fraction}{s}$ so this homogenous system determines moves that \emph{do not change the dimension of the solutions}.

Let's now consider the extended homogeneous system, where $\tilde{A}$ has already been defined in Section \ref{sec:gof},
\[
\tilde{A} \tilde{X} =0
\]  
with $\tilde{X}=(\tilde{x}_\zeta), \zeta \in \design$, $\tilde{x}_\zeta \in \mathbb Z$ and $\tilde{y}_\zeta+\tilde{x}_\zeta \geq 0$ for all $\zeta \in \design$.

Given $\tilde{Y}=(Y,\lambda_Y)$, where $Y$ is $\mathcal C$-compatible counting function and $\lambda_Y=\frac{\sum_\zeta y_\zeta}{s}$, the solutions of $\tilde{A} \tilde{X} =0$ determine all the other $\tilde{Y}+\tilde{X}=(Y+X,\lambda_{Y+X})$ such that $\tilde{A} (\tilde{Y}+\tilde{X}) =0$. $Y+X$ are $\mathcal C$-compatible counting functions whose sizes, $s \lambda_{Y+X}$, are, in general, \emph{different from that of $Y$}.

\subsection{Markov Basis}
We use the theory of Markov basis (see for example \cite{Drton|Sturmfels|Sullivant:2009} where it is also available a rich bibliography on this subject) to determine a set of generators of the moves. 
 
We use the following procedure in order to randomly select a $\mathcal C$-compatible counting function. We compute a Markov basis of $\ker(A)$ using 4ti2 (\cite{4ti2_1.3.1}). Once we have determined the Markov basis of $\ker(A)$, we make a random walk on the \emph{fiber} of $Y$, where $Y$, as usual, contains the values of the counting function of an initial design $\fraction$. The fiber is made by all the $\mathcal C$-compatible counting functions that have the same size of $\fraction$. The randow walk is done randomly choosing one move among the feasible ones, i.e. among the moves for which we do not get negative values for the new counting function. 

In the next paragraphs we consider moves for the cases that we have already studied in Section \ref{sec:gof}.
%\begin{remark}
%We finally observe that we got a homogeneous system thanks to the dimension preserving condition. If, given %$R=\sum_{p \in \design} y_p 1_p$, we want to explore fractions with different sizes it is enough that
%\begin{itemize}
%\item if we want to increase the dimension, we move to $kR=\sum_{p \in \design} k y_p 1_p$, where $k \in \Z, k %$>0$; we obtain a fraction with $k \cdot (\# \mathcal F)$ points
%$\item  if we want to decrease the dimension, we move to $\frac{1}{k}R=\sum_{p \in \design} \frac{1}{k} y_p 1_p$, %$where $ k \in \Z, 1 \leq k \leq \gcd(\{y_p \neq 0, p \in \design\})$ and $k$ divides all the $y_p \neq 0, p \in %$\design$; we obtain a fraction with $ \frac{1}{k} \cdot (\# \mathcal F)$ points
%$\end{itemize}
%$\end{remark}

\subsection{Orthogonal arrays}

\subsubsection{$OA(n,2^5,2)$}
We use the matrix $A$, already built in Section \ref{subsec:oa2_5} and give it as input to 4ti2 to obtain the Markov Basis, that we denote by $\mathcal{M}$. It contains $5.538$ different moves. Given $M=(x_\zeta)\in \mathcal M$ we define $M^+=\max(x_\zeta,0)$ and $M^{-}=\max(-x_\zeta,0)$. We have $M=M^+-M^-$. 

As an initial fraction $\fraction_0$, we consider the eight-run regular fraction whose indicator function $R_0$ is
\[
R_0=\frac{1}{4}(1+X_1X_2X_3)(1+X_1X_4X_5)
\]
We obtain the set of feasible moves observing that a move $M \in \mathcal{M}$, to be feasible, should be not negative when $R_0$ is equal to zero that means
\[
(1-R_0) M^{-} =0
\]
We find $12$ moves. Analogously an element $M \in \mathcal{M}$ such that 
\[
(1-R_0) M^{+} =0
\]
gives a feasible move, $-M$. In this case we do not find any of such element.

Therefore, given $R_0$, the set of feasible moves becomes $\mathcal M_{R_0}$ that contains $12+0$ different moves.

We randomly choose one move $M_{R_0}$ out of the $12$ available ones and move to
\[
R_1=R_0+M_{R_0}
\]
We run 1.000 simulations repeating the same loop, generating $R_i$ as $R_i=R_{i-1}+M_{R_{i-1}}$.

We obtain all the $60$ different 8-run fractions, each one with 8 different points as in \cite{carlini|pistone:2007}.

Using $\tilde{A}$ we obtain the set $\mathcal{\tilde{M}}$ that contains $18$ different moves. 
% and that is given in the Annex.

\subsubsection{$OA(n,3^3,2)$}
Using $A$ as built in the Section \ref{subsec:oa3_3}, we use 4ti2 to generate the Markov basis corresponding to the homogeneous system $AX=0$. We obtain $\mathcal{M}$ that contains $81$ different moves. 

As an initial fraction we can consider the nine-run regular fraction $\fraction_0$ whose indicator function $R_0$ is
\[
R_0=\frac{1}{3}(1+X_1 X_2 X_3+X_1^2 X_2^2 X_3^2)
\]
We run $1.000$ simulations repeating the same loop, i.e. generating $R_i$ as $R_i=R_{i-1}+M_{R_{i-1}}$.

We obtain all the $12$ different 9-run fractions, each one with 9 different points as known in the literature and as found in Section \ref{subsec:oa3_3}.

Using $\tilde{A}$ we also obtain the set $\mathcal{\tilde{M}}$ that contains $10$ different moves.
% (see Annex). 

\subsubsection{ $4 \times 4$ sudoku}
Using the matrix $A$ built in Section \ref{subsec:sudo4_4}, we run 4ti2 getting the Markov basis $\mathcal{M}$ that contains $34.920$ moves.

We randomly choose an initial sudoku

\renewcommand{\arraystretch}{1.2}
  \begin{equation*} \begin{array} { |c c|c c|} \cline{1-4}
   3 &   2  &  4  &  1 \\
   4 &   1  &  3  &  2 \\ \cline{1-4}
   2 &   3  &  1  &  4  \\
   1 &   4  &  2  &  3 \\ \cline{1-4}
   \end{array}
\end{equation*}
The corresponding indicator function is
\[
F_0=\frac 1 4 ( 1 - R_2 C_1 S_1 S_2 ) ( 1 - R_1 C_2 S_1) \ .
\]  
Then we extract from $\mathcal{M}$ the feasible moves. We obtain a subset $\mathcal M_{F_0}$ that contains $5$ different moves. We repeat the procedure on $-\mathcal{M}$ and we obtain other $9$ moves. 
 
We randomly choose one move $M_{F_0}$ out of the $5+9$ available ones and move to
\[
F_1=F_0+M_{F_0}
\]
We run $1.000$ simulations repeating the same loop $F_i=F_{i-1}+M_{F_{i-1}}$.

We obtained all the $288$ different $4 \times 4$ sudoku.  

\section{Conclusions}
We considered mixed level fractional factorial designs.  Given the counting function $R$ of a fraction $\fraction$ we translated the constraint $c_\alpha=0$, where $c_\alpha$ is a generic coefficient of its polynomial representation $R=\sum_{\alpha}c_{\alpha} X^{\alpha}$, into a set of linear constraints with integer coefficients on the values $y_\zeta$ that $R$ takes on all the points $\zeta \in \design$. We obtained the set of generators of the solutions of some problems using Hilbert Basis. We also studied the moves between fractions. We characterized these moves as the solution of a homogeneous linear system. We defined a procedure to randomly walk among the solutions that is based on the Markov basis of this system. We showed the procedure on some examples. Computations have been made using 4ti2 (\cite{4ti2_1.3.1}). 

Main advantages of the procedure are that we do not put restrictions on the number of levels of factors and that it is not necessary to use software that deals with complex polynomials.  

One limit is in the high computational effort that is required. In particular only a small part of the Markov basis is used because of the requirement that counting functions can only take values greater than or equal to zero. The possibility to generate only the moves that are feasible could make the entire process more efficient and is part of current research.
% \appendix

% \section{}
% \label{}

%\bibliographystyle{elsart-harv}
%\bibliography{C:/tuttodae07}
%\bibliography{/Users/gianni/Archive/bibs/tutto}
%\bibliography{C:/tuttomp}
%\bibliography{C:/tutto_cineca}

\end{document}